\documentclass[a4paper,11pt]{article}
\usepackage{pos}
\usepackage{wrapfig}

\title{Selected topics on 
the QCD phase diagram at finite temperature and density}

\author*[a]{Christian Schmidt}

\affiliation[a]{Fakultät für Physik, Universität Bielefeld, Universitätsstraße 25, 33615 Bielefeld, Germany}

\emailAdd{schmidt@physik.uni-bielefeld.de}

\abstract{We will report recent progress on the QCD phase diagram at finite temperature and density. In particular, we discuss the universal scaling of the chiral transition in the limit of two massless quarks and one strange quark. We also discuss influence of other control parameter as chemical potentials, external magnetic field strength and number of quark flavors on the chiral transition. From calculations of Taylor expansion coefficients of the pressure w.r.t the baryon chemical potential and at imaginary chemical potential, we discuss estimates of the QCD critical point. Those estimates make use of the universal scaling ansatz of the Lee-Yang edge singularity.  }

\FullConference{The 41st International Symposium on Lattice Field Theory (LATTICE2024)\\
 28 July - 3 August 2024\\
Liverpool, UK\\}

\begin{document}
\maketitle
\section{Introduction}
\begin{wrapfigure}{r}{0.5\textwidth}
    \centering
    \includegraphics[width=0.4\textwidth]{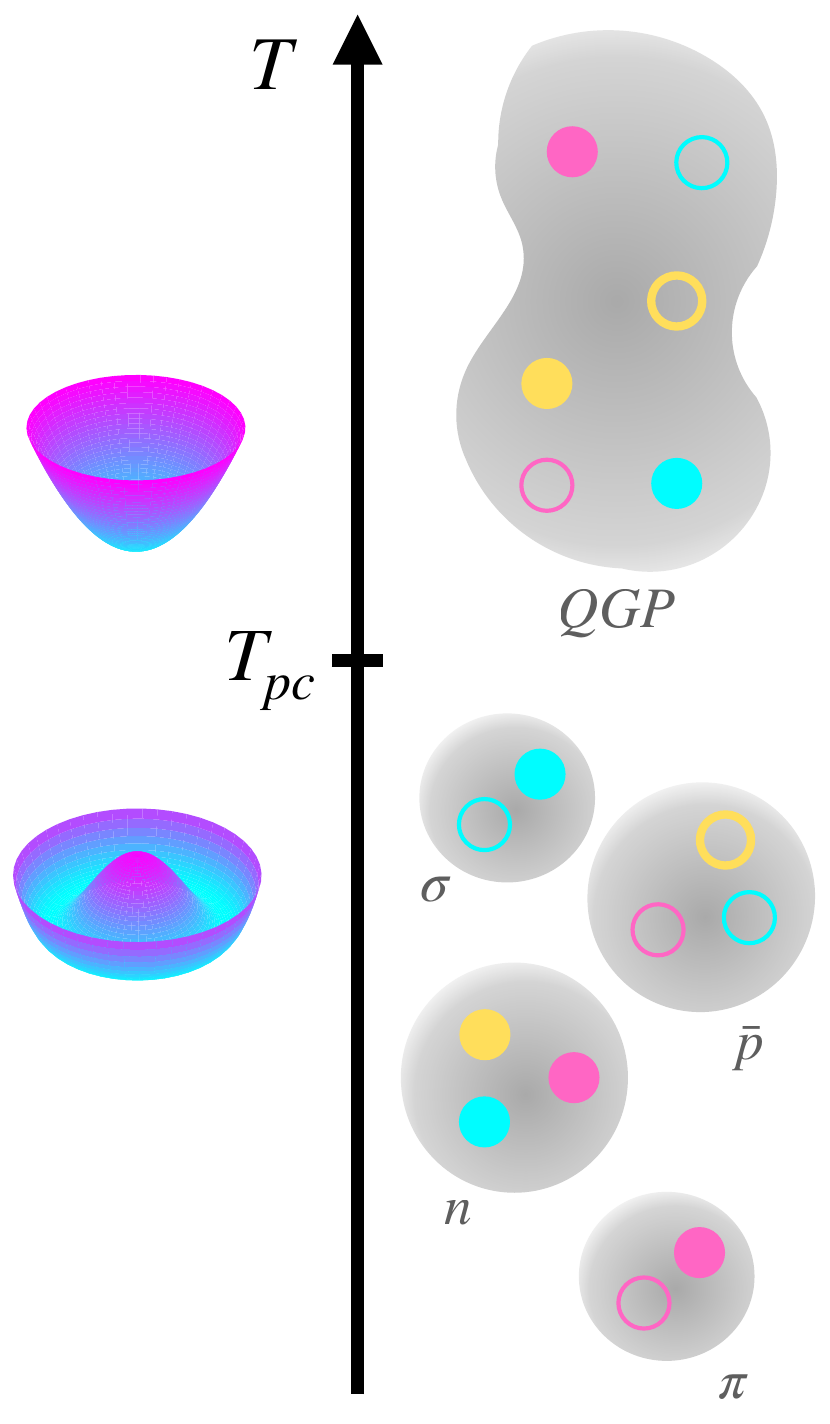}
    \caption{Schematic representation of the chiral and deconfinement transition in QCD. }
    \label{fig:chiral_vs_deconfinement}
\end{wrapfigure}
The theory of strong interaction matter as described by Quantum Chromodynamics (QCD),  exhibits two important nonperturbative features, spontaneous breaking of the chiral symmetry and color confinement. 
In fact, both phenomena are related to symmetries of the theory. 
In the former case, the chiral symmetry denotes the invariance of the action under independent unitary rotations of left- and right-handed Dirac spinors in flavor space, $\text{SU}(N_f)_\text{L}$ $\times$  $\text{SU}(N_f)_\text{R}$. 
This symmetry is manifest in the massless Lagrangian but explicitly broken by the quark mass term and, in addition, spontaneously broken at low temperatures.
The latter phenomena is related to the limit of infinitely heavy quarks (quenched limit of QCD), where the action is invariant under generalized gauge transformations w.r.t. the center of the gauge group, Z(3). While the chiral transition can be analysed in terms of an effective Hamiltonian of an order parameter field, the chiral condensate \cite{Pisarski:1983ms}, the mechanism of color confinement is still not very well understood. One physical picture is obtained in terms of the Polyakov loop, which is related to the free energy of a static color charge as probe of the system. For a review on Polyakov loop modeling for hot QCD see, e.g. \cite{Fukushima:2017csk}. 
It is a nontrivial observation that the chiral and deconfinement transition temperatures are in good agreement, at least at vanishing baryon number density \cite{Karsch:1994hm, Bazavov:2011nk}. 
We note, however, that the notion of these temperatures is only uniquely defined through the order parameters chiral condensate and Polyakov loop in the respective limits of massless and infinitely heavy quarks. 
At any small but nonzero quark mass, the order of the transition is a crossover \cite{Aoki:2006we} and each observables might define a different (possibly pseudo-critical) temperature $T_{pc}$.
A schematic picture is shown in Fig.~\ref{fig:chiral_vs_deconfinement}. 
On the left side the chiral potential is depicted, which gives rise to a nonvanishing vacuum expectation value of the chiral condensate below $T_{pc}$. 
On the right side the perculation picture of the deconfinement transition is represented, where color charges can move over large distances once hadrons have sufficient overlap above $T_{pc}$. It is interesting to mention that the role of the Polyakov loop changes from an order parameter (magnetization-like operator) at very large masses to a energy-like operator near the chiral limit \cite{Clarke:2020htu}.
The interplay between chiral and deconfinement transition might also be discussed in terms of the spectrum of the Dirac operator. 
While the chiral condensate is very sensitive to the low eigenvalues \cite{Banks:1979yr}, the Polyakov loop is more sensitive to the bulk of the spectrum \cite{Bruckmann:2006kx}. 
Additional information is encoded in the degree of localization of eigenmodes. The transition between localized and unlocalized eigenmodes (Anderson transition) might be another manifestation of deconfinement \cite{Kehr:2023wrs, Kehr:2025nth}. 

In the following, we will briefly motivate additional control parameters, that influence the QCD transition (Sec.~\ref{sec:pdiag}) before we analyse the universal critical scaling near the the chiral transition in more detail (Sec.~\ref{sec:scaling}). We will argue that scaling fits are a powerful tool to extract important nonuniversal constants such as the chiral transition temperature from lattice data. In Sec.~\ref{sec:charges} we will comeback to the deconfinement transition, which we will assess in terms of conserved charge fluctuations and the melting of hadronic states. Finally we will use the same type of observables and to constrain the location of the QCD critical point (Sec.~\ref{sec:CEP}).

\section{A walk through the QCD phase diagram\label{sec:pdiag}}
\begin{figure}
    \centering
    \includegraphics[width=0.7\textwidth]{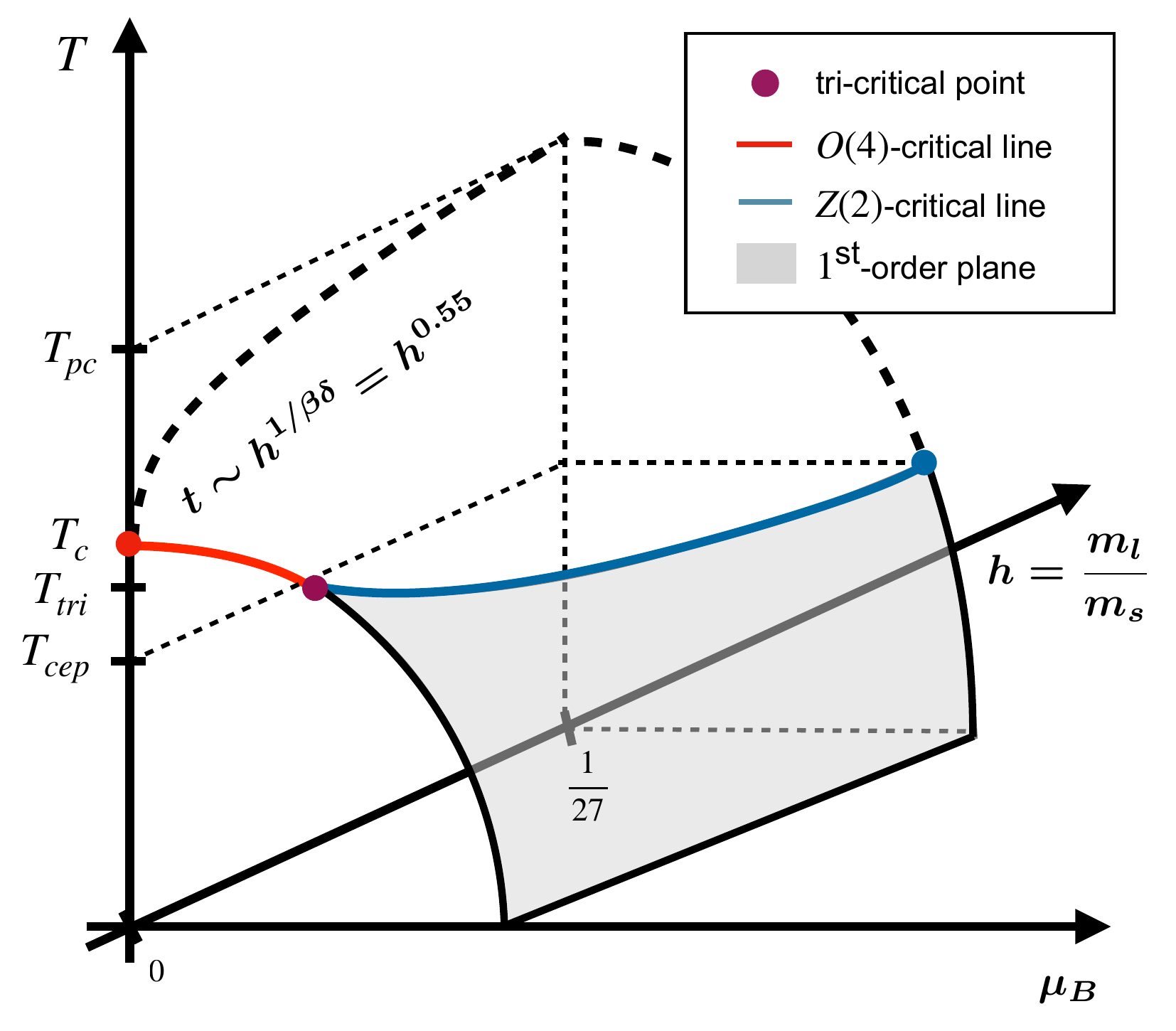}
    \caption{Schematic view of the (2+1)-flavor phase diagram in the parameter space spanned by temperature $T$, quark mass ratio $m_l/m_s$ and baryon chemical potential $\mu_B$.}
    \label{fig:pdiag}
\end{figure}

A schematic view of the (2+1)-flavor phase diagram is indicated in Fig.~\ref{fig:pdiag}. 
Our current understanding of this phase diagram assumes a second order phase transition in the chiral limit. Despite the fact that the symmetry breaking pattern of the chiral transition, and thus also its nature, is not established beyond any doubt \cite{Pelissetto:2013hqa}, we assume here that the axial anomaly $\text{U}(1)_\text{A}$ remains broken at $T_c$ \cite{Kovacs:2025wdl}, while the chiral symmetry breaks such that $\text{SU}(N_f)_\text{L}$ $\times$  $\text{SU}(N_f)_\text{R}$ $\to$ $\text{SU}(N_f)_\text{V}$ for $T<T_c$. With two degenerate light quark flavors (up and down), this would correspond to a second order transition in the universality class of the $O(4)$ spin-model. In (2+1)-flavor QCD, the light quark mass $m_l$ is thus an important control parameter that takes over the role of the symmetry breaking field $h$. It is convenient to define the two scaling fields $t,h$ as 
\begin{equation}
    t=t_0^{-1}(T/T_c-1) \quad \text{and}\quad h=h_0^{-1}(m_l/m_s)\;.
\end{equation}
Here $m_s$ denotes the strange quark mass and $t_0, h_0$ are nonuniversal normalization constants that need to be determined. 
The same is true for the chiral transition temperature $T_c$. The traditional method to determine $T_c$ is by locating the peak position of the chiral susceptibility $\chi_l$, as a function of $h$, which defines one possible pseudo-critical line. A similar method was pursued in Ref.~\cite{HotQCD:2019xnw}, where the position that corresponds to 60\% of the maximum of the peak height (for $T<T_c$) was determined. A chiral critical temperature of $T_c=132^{+3}_{-6}$ MeV \cite{HotQCD:2019xnw} was found, which is considerably smaller than the crossover temperature $T_{pc}=156.5(1.5)$ MeV at physical quark masses \cite{HotQCD:2018pds}. In the vicinity of the chiral critical point, indicated as red dot in Fig.~\ref{fig:pdiag}, the pseudo-critical line is given as $t\sim h^{1/\beta\delta}$, with $\beta,\delta$ being critical exponents.

Already since the introduction of QCD, people though about the dependence of the chiral transition on the baryon number density and its connection with the onset of matter. 
Due to the charge conjugation symmetry, and since the net baryon chemical potential $\mu_B$ does not break the chiral symmetry, the parameter can be added to the reduced temperature as 
\begin{equation}
    t=t_0^{-1}((T/T_c-1)+\kappa_2^B\hat\mu_B^2)\;,
\end{equation}
where we denote $\hat\mu_B=\mu_B/T$. Lattice QCD calculations are hindered by the infamous sign problem. 
At small chemical potentials the sign problem can be overcome by a Taylor expansion approach \cite{Allton:2002zi}. The curvature coefficient $\kappa_2^B$ can be determined by expanding thermodynamic quantities in $\hat\mu_B$ about $\hat\mu_B=0$. 
It was found that the value depends only very mildly on the light quark mass \cite{Ding:2024sux}. 
The value reduces by about 10\% under the constraint of a vanishing strange quark density ($n_S=0$) \cite{Ding:2024sux}. 
In this case, the continuum extrapolated result is $\kappa_2^B=0.012(4)$ \cite{HotQCD:2018pds}, a consistent results of $\kappa_2^B=0.0153(18)$ was given in Ref.~\cite{Borsanyi:2020fev}. As a consequence, $T_{pc}(\mu_B)$ and $T_c(\mu_B)$ are decreasing with increasing $\mu_B$, as indicated in Fig.~\ref{fig:pdiag} by the dashed and read lines. 

At larger chemical potentials people expect a second order phase transition point as an endpoint of a first order line, known as the QCD critical end-point (CEP). 
Due to the emergent discrete reflection symmetry (Z(2)) in the vicinity of the CEP, it will be in the universality class of the 3d-Ising model.
It has been considered a grand challenge to find this point by lattice QCD methods. Current advances in this direction will be discussed in Sec.~\ref{sec:CEP}. 
It is interesting to mention that some recent works from lattice, Functional Renormalization Group (FRG) and holographic models find a critical point in a similar region at $(T_{cep},\mu_{B,cep})\approx(110,600)$ MeV. 
Considering the light quark mass dependence of the QCD critical point, a line of second order phase transitions is formed, indicated as blue line in Fig.~\ref{fig:pdiag}. 
The line originates from a tri-critical point ($T_{tri}$) in the chiral limit, indicated as a purple dot. 
In the vicinity of the tri-critical point, its behavior is gouverned by tri-critical exponents. 
It thus emerges a hierarchy of important temperatures: $T_{pc}>T_c>T_{tri}>T_{cep}$.

\begin{figure}
    \centering
    \includegraphics[width=0.4\textwidth]{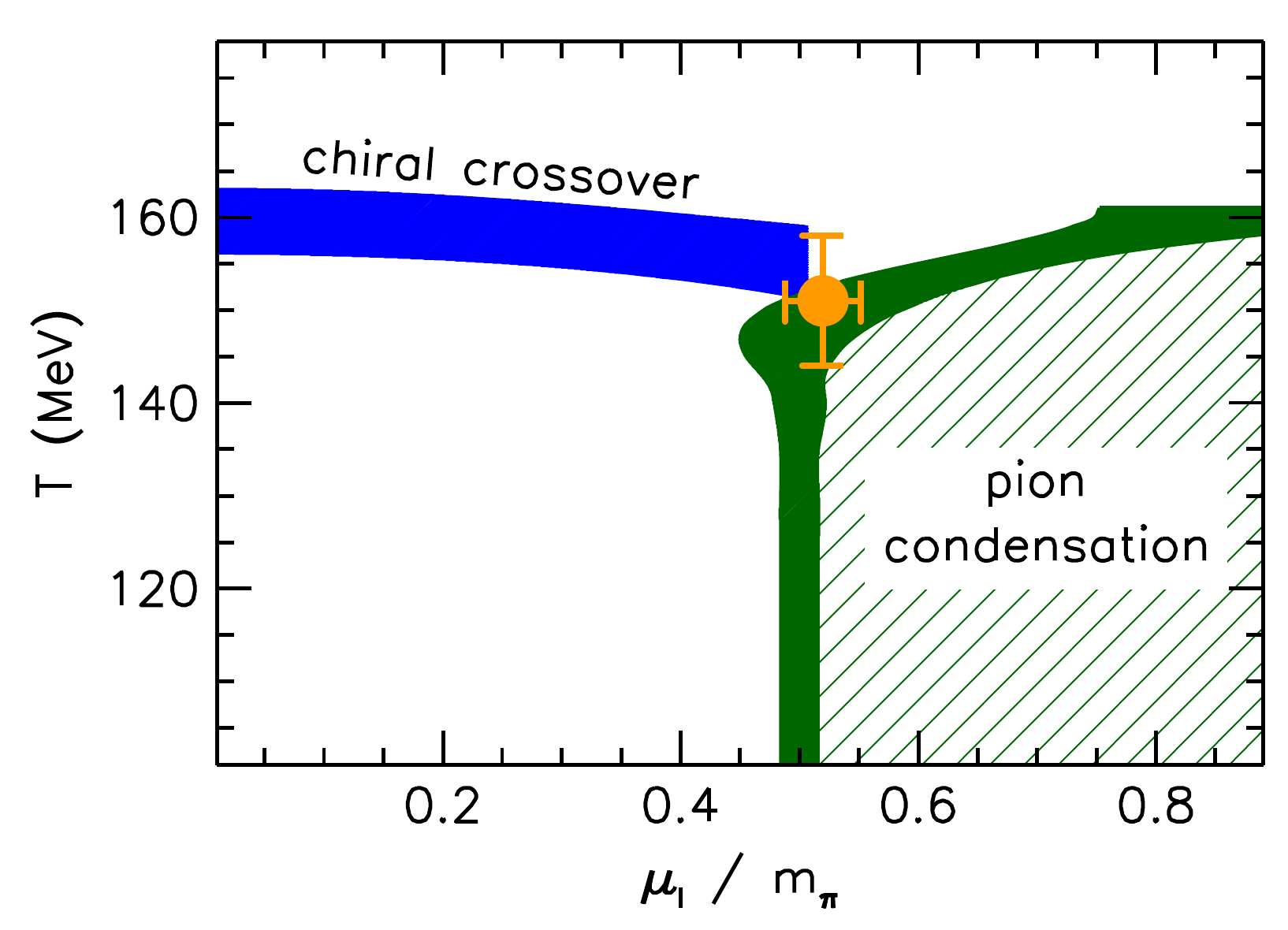}
    \includegraphics[width=0.55\textwidth]{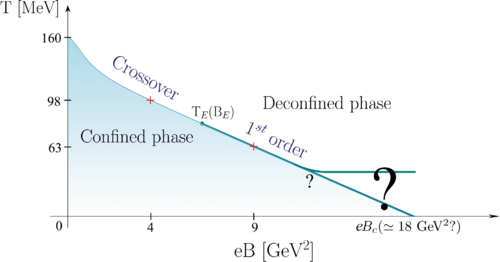}\\
    \includegraphics[width=0.95\textwidth]{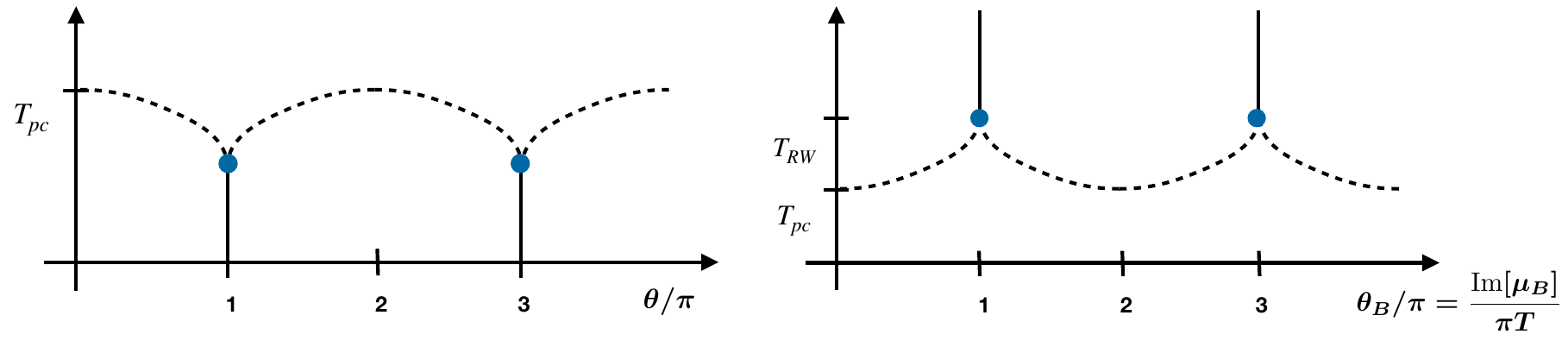}
    \caption{Different versions of the QCD phase diagram with temperature and one further control parameter, shown are isospin chemical potential $\mu_I$ (top left), exteranal magnetic field $eB$ (top right), topological angle $\theta$ (bottom left) and imaginary chemical potential $\theta_B$ (bottom right). The top diagrams are taken from Ref.~\cite{Brandt:2017oyy}, and Ref.~\cite{DElia:2021yvk}, respectively.}
    \label{fig:pdiag_other}
\end{figure}

It is interesting and phenomenological important to explore also other control parameter in the QCD phase diagram, as shown in Fig.~\ref{fig:pdiag_other}.
An obvious extension to the list of control parameter are further chemical potentials. 
In (2+1)-flavor QCD there are three independent chemical potentials (one for each quark flavor), which can be transformed into the corresponding chemical potentials of the hadronic charges, such as baryon number ($B$), strangeness ($S$), and electric charge ($Q$). Often one also discusses isospin ($I$). A nonzero $\hat\mu_I$ gives rise to a charged pion condensate for $\mu_I>m_\pi/2$. The corresponding phase diagram is shown in Fig.~\ref{fig:pdiag_other} (top right) \cite{Brandt:2017oyy}, which is of interest for at least two reasons: (i) a pion condensation phase in the early Universe might have been triggered by a large lepton asymmetry \cite{Middeldorf-Wygas:2020glx, Vovchenko:2020crk}, and (ii) a class of compact stars which is mainly composed out of a Bose-Einstein condensate of charged pions \cite{Brandt:2018bwq} might exist. Even though the lattice computations at $\mu_I>0$ do not suffer from a sign problem as in the $\mu_B>0$ case, they are still not completely trivial. The calculation have to be performed with a isospin-breaking term present, which relative strength has to be extrapolated to zero \cite{Kogut:2002zg}. A improved method for the extrapolation has been introduced in \cite{Brandt:2017oyy}, see also \cite{Basta:2025svw, Brandt:2025hpy}. A completely different method relies on the calculation of n-point correlation function of charged pions, which can be pushed to n as large as 6144 \cite{Abbott:2023coj}.

A further control parameter which is often discussed is the external magnetic field $eB$, Fig.~\ref{fig:pdiag_other} (top right). The interest into strong magnetic fields is again motivated by the (i) phenomenology of heavy ion collisions (ii) compact stars and (iii) the early universe, where in all cases strong magnetic fields can be generated. In particular, it was estimated that in peripheral heavy ion collisions at LHC energies magnetic fields of up to $eB\sim 15 m_\pi^2$ could be generated \cite{Skokov:2009qp}. It was observed that the pseudo critical temperature $T_{pc}(B)$ is decreasing with increasing magnetic field \cite{Bali:2011qj, Bali:2012zg, Endrodi:2015oba}, which is in contrast to analytic predictions \cite{Andersen:2014xxa}. At strong magnetic fields as large as $eB=9GeV^2$ a first order transition was observed \cite{DElia:2021yvk}, which is indicative for a critical point in the range of $eB=4$ -- $9GeV^2$. At asymptotically large magnetic fields, quarks might decouple from the theory \cite{Miransky:2002rp}, which raise the question whether the first order line terminates at $T=0$ or if it becomes constant in $eB$. This is indicated with a question mark in Fig.~\ref{fig:pdiag_other} (top right). For further influence of the external magnetic field on the QCD thermodynamics see also \cite{Brandt:2024fpc, Brandt:2025now} (chiral magnetic effect) \cite{Brandt:2024gso} (topological sysceptibility) \cite{MarquesValois:2025nzo, Kumar:2025ikm, Ding:2025siy} (equation of state and conserved charge fluctuations). 

Finally we mention that the dependence of the phase diagram on the topological vacuum angle $\theta$ is also studied \cite{DElia:2013uaf}, which is motivated by axion physics and the axial anomaly \cite{GrillidiCortona:2015jxo}. It was proposed that the $T-\theta$ phase diagram is analogous to the phase diagram wit imaginary chemical potential $\theta_B=\text{Im}[\mu_B]$ \cite{DElia:2013uaf}, see Fig.~\ref{fig:pdiag_other} (bottom). We note that direct simulations with $\theta>0$ are unfeasible because of the obvious sing problem in the gauge action. Just like in the case of $\mu_B$, the problem is circumvented by either performing a Taylor expansion in $\theta$ about $\theta=0$, or by performing simulations at imaginary $\theta$. 

\section{Universal scaling and chiral symmetry restoration\label{sec:scaling}}
\begin{figure}
    \centering
    \includegraphics[width=0.475\textwidth]{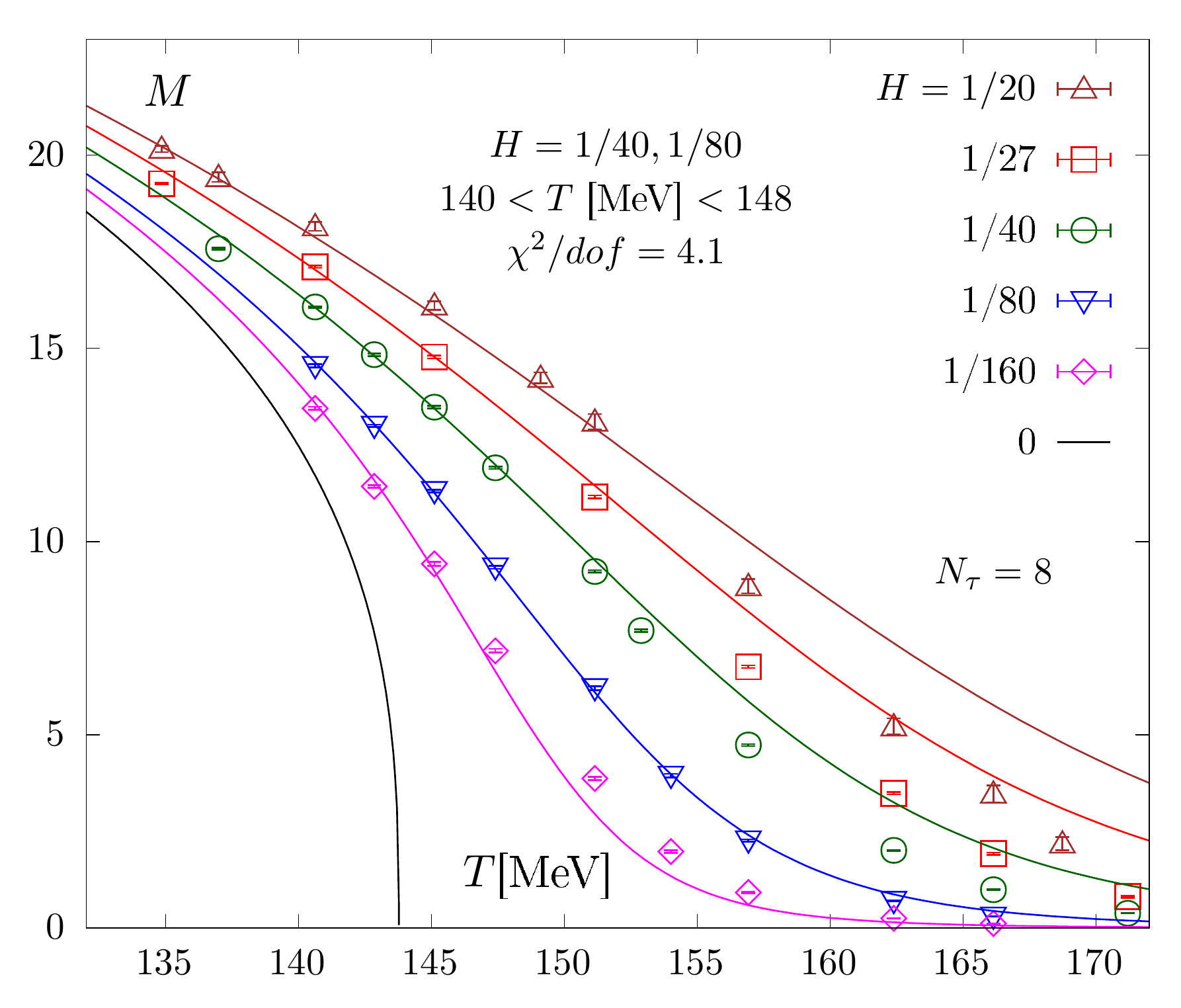}
    \raisebox{3pt}{\includegraphics[width=0.505\textwidth]{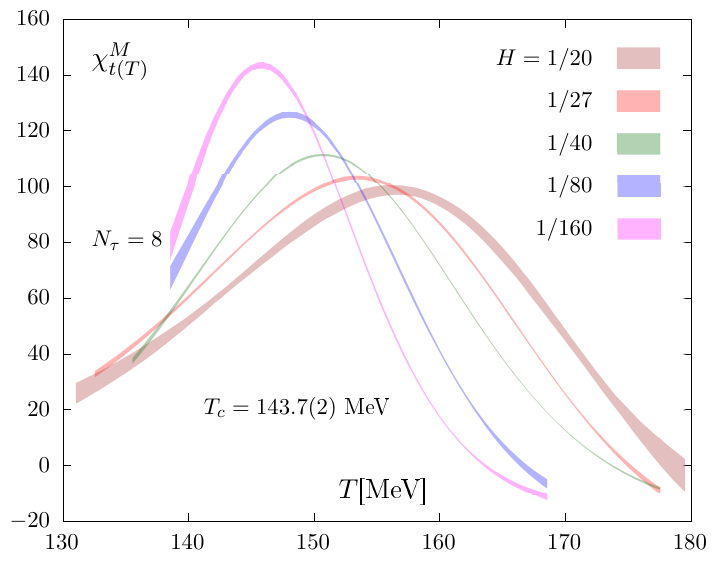}}
    \caption{Left: The dimensionless renormalized order parameter $M$ versus $T$. Shown also is a fit to the data for $H = 1/40$ and $1/80$ in the temperature interval $T \in [140 \text{MeV}, 148 \text{MeV}]$. Right: The temperature derivative of the improved order parameter, obtained from rational function fits to the data. }
    \label{fig:scaling_fit}
\end{figure}

It is very important to investigate to what extent universal scaling related to the chiral transition can be observed in lattice QCD calculations. Universal scaling assumes that the free energy has a singular and a regular part $f(T,H,L)=f_s(T,H,L)+f_{\text{reg}}(T,H,L)$, where the state variables $T,H,L$ are the temperature, the external field and the system size, respectively. We further assume that the singular part is a generalized homogeneous function of the reduced scaling fields $t,h,l$,
\begin{equation}
f_s(t,h,l)=b^{-d}f_s(b^{y_t}t,b^{y_h}h,bl)\;,
\end{equation}
which is the scaling hypothesis, motivated by the fact that there is no physical length scale near a critical point due to critical fluctuations. The exponents $y_t=1/\nu$  and $y_h=\beta\delta$, define the critical exponents $\beta,\delta,\nu$, which are related to each other through the hyperscaling relation $\delta=d\nu/\beta -1$. If we chose the (arbitrary) scale parameter $b$ to keep one of the scaling fields constant, we arrive at a scaling function with has one argument less. A popular choice is to chose $b=h^{-1\beta\delta}$, which will eliminate the $h$ dependence and introduces the scaling variable $z=t/h^{1\beta\delta}$. The standard order parameter and the magnetic susceptiblity are given as $M=-\partial f/\partial H$, $\chi_h=\partial M/\partial H$.
The corresponding scaling functions, 
\begin{equation}
    M(T,H,L)=h^{1/\delta}f_G(z,z_L) +\text{reg} \quad \text{and} \quad \chi_h=h_0^{-1}h^{1/\delta -1}f_\chi(z,z_L)+\text{reg}\;,
\end{equation}
are known for the relevant Universality classes Z(2), O(2) and O(4) \cite{Karsch:2023pga}. At this point we make contact with QCD observables. In QCD the (bare) ciral condensate and chiral susceptiblity are given as 
\begin{equation}
    M_l=\frac{m_s}{f_K^4}\frac{T}{V}\frac{\partial \ln Z}{\partial m_l}\,, \quad \chi_l=\frac{m_s^2}{f_K^4}\frac{T}{V}\frac{\partial^2 \ln Z}{\partial m_l^2}\,,
\end{equation}
where the factors $m_s$ ($m_s^2$) remove multiplicative divergences. The multiplication with an appropriate power of the kaon decay constant $f_K$ makes the observables dimensionless. The partial derivative with respect to the light quark mass is to be understood as $\partial/\partial m_l=\partial/\partial  m_u+\partial/\partial m_d$. The remaining UV divergences still need to be subtracted. In the past this was done by subtracting a fraction of the strange condensate \cite{Cheng:2007jq, Ejiri:2009ac}. However, that construction is not directly related to a scaling function. It is advantageous to introduce an improved order parameter \cite{Unger:2010wcq, Kotov:2021rah}, given as  
\begin{equation}
    M=M_l-H\chi_l=h^{1/\delta}(f_G(z,z_l)-f_\chi(z,z_L))\;.
    \label{eq:eos}
\end{equation}
This order parameter is renormalized, fulfills an equation of state in terms of scaling functions and also removes the leading contributions of the regular part.  In Fig.~\ref{fig:scaling_fit} (left)
results for the improved order parameter from highly improved staggered quarks (HISQ) are shown \cite{Ding:2024sux}. The calculations are from lattices with temporal lattice extent $N_\tau=8$ and the spatial extent $N_\sigma$, which has been increased with decreasing $m_l$, i.e.
$4 \le N_\sigma \le 7$, insuring that the inverse of the pion correlation length stays approximately constant, $m_\pi L \sim (3-4)$. Also shown is a fit to the equation of state Eq.~(\ref{eq:eos}). The overall agreement is quite satisfactory, however, to obtain a good $\chi^2$, the temperature interval has to be reduced to $T \in [140 \text{MeV}, 148 \text{MeV}]$ and the quark mass ratio is restricted to $H\in[1/80, 1/40]$. The scaling function and critical exponents are those of the O(2) universality class. The reason is that staggered fermions break the chiral symmetry such that the expected O(4) symmetry for two light flavors is further reduced to an O(2) symmetry for any fixed lattice spacing. The fit yields a critical temperature of $T_c=143.7(2)$ MeV, which is in agreement with the corresponding $N_\tau=8$ value from \cite{HotQCD:2019xnw}.

The size of the scaling region is controversial. While FRG studies predict a small scaling region which requires $m_\pi \lesssim (2-5)$ MeV \cite{Braun:2023qak}, tt was found to reach out to the physical point in Ref.~\cite{Kotov:2021rah}. We note, however, that the notion of scaling region is not unique and strongly depends on the precise definition. It is supposed to be a measure on how much the physics at the physical point is determined by universal behaviour. 

\begin{wrapfigure}{r}{0.5\textwidth}    
    \centering
    \includegraphics[width=0.5\textwidth]{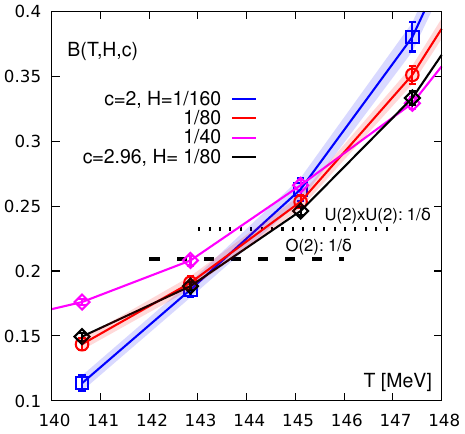}
    \caption{The logarithm of ratios of the improved order parameter versus temperature as introduce in Eq.~(\ref{eq:delta}). The dashed and dotted lines give the values for $1/\delta$ in the O(2) and U(2) $\times U(2)$ \cite{Pelissetto:2013hqa} universality classes, respectively.}
    \label{fig:universality}
\end{wrapfigure}

In order to avoid the assumption of the universality class, which is necessary to perform a scaling fit, one can also plot normalized ratios of the improved order parameter end extract the value of the critical exponent delta, in particular from Eq.~(\ref{eq:eos}) and the known normalization of the scaling functions we can derive 
\begin{equation}
    M(T_c,H)/H^{1/\delta}=h_0^{-1/\delta}(1-1/\delta) \quad \text{and} \quad
    \ln\left(\frac{M(T_c,H_1)/H_1^{1/\delta}}{M(T_c,H_2)/H_2^{1/\delta}}\right)=\frac{1}{\delta}\ln(H_1/H_2)\;.
    \label{eq:delta}
\end{equation}
While the first equation shows that $M(T,H)/H^{1/\delta}$ has a unique crossing point at $T=T_c$, which can be used to determine $T_c$ without any scaling fit, the second equation can be used to determine $\delta$ and thus the universality class. Preliminary results are shown in Fig.~\ref{fig:universality}, using $c=H_1/H_2=2$ and $c=2.96$.

While the data is still lacking sufficient precision to further resolve the unique crossing point, the value of $\delta$ currently favours the O(2) universality class, which corresponds to the O(4) universality class in the continuum. This would indicate that the axial anomaly is still present at $T=T_c$. The alternative universality class U(2) $\times$ U(2), is disfavoured. The values of $\delta$ for both universality classes are indicated as dashed lines. 

Results presented so far are based on (2+1)-flavor calculations with staggered fermions. It will be important to verify universal scaling also with other lattice discretizations. In Ref.~\cite{Kotov:2021rah}, twisted mass Wilson fermions were used. Universal scaling was found that is compatible with O(4) scaling. The estimated chiral transition temperature $T_c=134^{+6}_{-4}$ MeV is in good agreement with the HotQCD result $T_c=134^{+2}_{-6}$ MeV \cite{HotQCD:2019xnw}. Calculations with Möbius domain wall fermions at finite temperature are also underway \cite{Zhang:2025vns, Goswami:2025euh, Ward:2024wze}. A preliminary pseuo-critical transition temperature of $T_{pc}=153(2)$ MeV was found on $36^3\times 12$ configurations, which is consistent with the HotQCD ($T_{pc}=156.5(1.5)$ MeV) \cite{HotQCD:2018pds} and Wuppertal-Budapest ($T_{pc}=158.0(0.6)$ MeV) \cite{Borsanyi:2020fev} results. 

\begin{figure}
    \centering
    \includegraphics[width=0.42\textwidth]{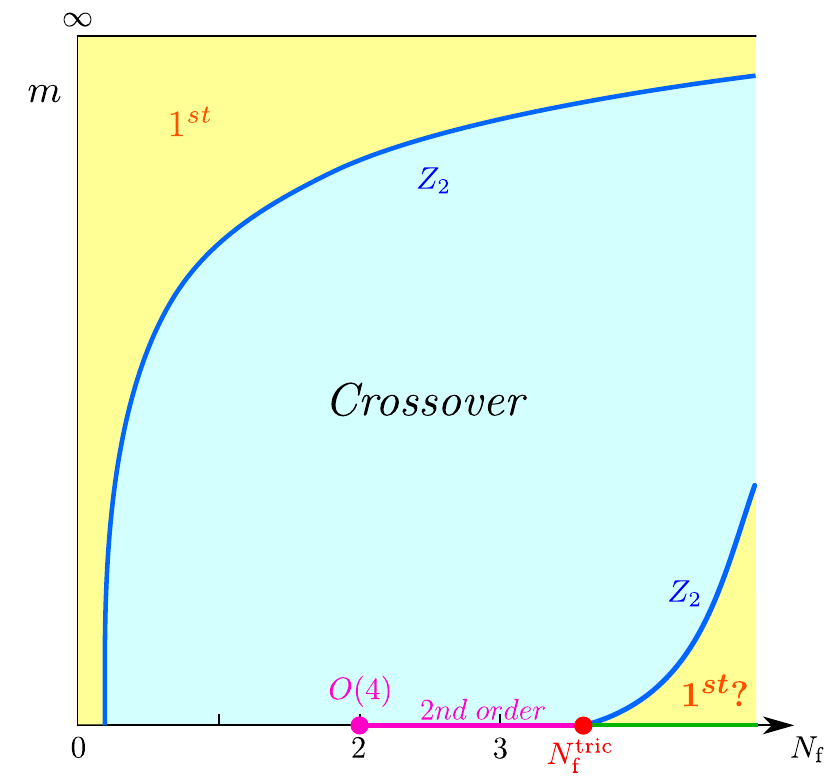}
    \includegraphics[width=0.56\textwidth]{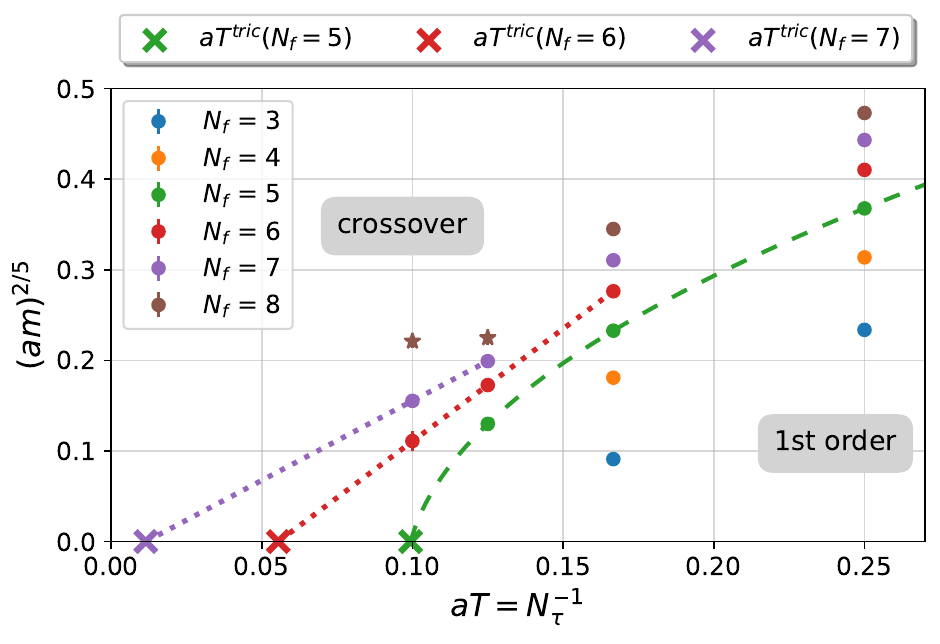}

    \caption{Left: Diagram that indicates the order of the QCD transition in the $m,N_f$ plane. The figure was taken from \cite{Cuteri:2021ikv}. Right: critical temperature on the Z(2) boundary line between crossover and first order region. The fits are motivated by the expected tricritical scaling. The figure was taken from \cite{Klinger:2025xxb}.}
    \label{fig:nf}
\end{figure}

The control parameter which we have neglected so far is the number of light quark flavors $N_f$. The seminal work of Pisarki and Wilczek \cite{Pisarski:1983ms} predicted a first-order chiral phase transition for $N_f\ge3$, i.e. QCD with three mass-degenerate quark flavors. This statement was not verified in lattice calculations. Albeit first-order transitions have been found on coarse lattices \cite{Karsch:2001nf, deForcrand:2003vyj}, it turned out that the first order region for $N_f=3$ is strongly cut-off dependent and likely vanishes in the continuum limit.  Calculations with HISQ fermions at $N_\tau=6,8$ found evidence for a second order critical point in the chiral limit \cite{Bazavov:2017xul, Dini:2021hug}, as in the (2+1)-flavor case \cite{Ding:2024sux}. The chiral transition temperature $T_c(N_f=3)$ was estimated to be $98^{3}_{-6}$ MeV. Although a very small critical quark mass can not be excluded, this seems currently unlikely. In calculations with Möbius domain wall fermions no direct evidence for a first order region could be identified \cite{Zhang:2025vns}. The difficulties in the determination of a first order region for $N_f=3$ motivated calculations with $N_f\ge3$, where the strength of the first order region is predicted to be larger. A rigorous scan of the bare parameter space $\{\beta,am,N_f,N_\sigma,N_\tau\}$ with staggered fermions was conducted \cite{Cuteri:2017gci,Cuteri:2021ikv,Klinger:2025xxb}. The problem was also generalized to non integer numbers of flavors in order to find the tri-critical point $N_f^{tric}$ in the version of the Columbia plot shown in Fig.~\ref{fig:nf} (right). The tri-critical point indicates the flavor number where the order of the transition in the chiral limit changes from second order to first order. 
The strategy is the following: In the first step the critical coupling $\beta_c$ and the critical quark mass $am_c$ are determined for each of the combinations $\{N_f,N_\tau\}$. This is done by a finite size scaling analysis of the kurtosis of the order parameter 
\begin{equation}
    B_4=\frac{\left<(M-\left<M\right>)^4\right>}{\left<(M-\left<M\right>)^2\right>^2}
\end{equation} 
which is also known as Binder's Cumulant \cite{PhysRevLett.47.693}. This quantity has a universal, volume independent value at the critical point. The expected Z(2) value is $B_4=1.604$ and the Z(2) finite size scaling is found to very good precision \cite{Karsch:2001nf}. All the critical points $(\beta_c,am_c)$ are located on the Z(2)-critical line that originates from the tri-citical point as shown in Fig.~\ref{fig:nf} (right). The second step is now the extrapolation of the Z(2) critical points to the tri-critical point in the chiral limit. This can be done by exploiting the known tri-critical scaling at fixed $N_\tau$ or at fixed $N_f$. For the critical temperature we have
\begin{equation}
    T_c(m)=T_{tric}+Am^{2/5}+Bm^{4/5}+\mathcal{O}(m^{6/5})\;.
\end{equation}
The extrapolation at fixed $N_f$ in lattice units is shown in Fig.~\ref{fig:nf} (right). We note that the first order region below the curve only survives the continuum extrapolation, if the curve is connected with the origin, representing the continuum limit. The conclusion is that the tri-citical point is located at $N_f^{tric}>7$. For all $N_f\le7$, the observed first order region is a lattice artefact. It is interesting to mention that the same analysis was performed on the published data from $\mathcal{O}(a)$ improved Wilson calculations at $N_f=3$ \cite{Kuramashi:2020meg} with the same conclusion, i.e. also the observed first order region in the Wilson calculations vanish in the continuum. The apparent discrepancies between the lattice calculations and Ref.~\cite{Pisarski:1983ms} was recently re-investigated in the linear sigma model and Ginzburg-Landau models \cite{Pisarski:2024esv,Giacosa:2024orp,Fejos:2024bgl}. The discrepancy was not resolved but the possibility of a undetectable small critical mass was pointed out. 

Given the above findings, we can now ask if the chiral transition temperature $T_c(N_f)$ remains finite at $N_f=N_f^{tric}$ or not. If $T_c(N_f^{tri})=0$, we hit a quantum tricitical point and enter the conformal window. 
Current estimates for the lower bound of the conformal window $N_f^*$ suggest $8\lesssim N_f^*\lesssim 12$ \cite{Braun:2009ns, Hasenfratz:2023wbr, Hasenfratz:2024fad, Deuzeman:2009mh, Kotov:2021hri, Miura:2011mc, Miura:2012zqa, Braun:2006jd}. This does not leave much room for the existence of a first order region below the conformal window.

\section{Conserved charge fluctuations and effective degrees of freedom\label{sec:charges}}
\begin{figure}
    \centering
    \raisebox{7pt}{\includegraphics[width=0.517\textwidth]{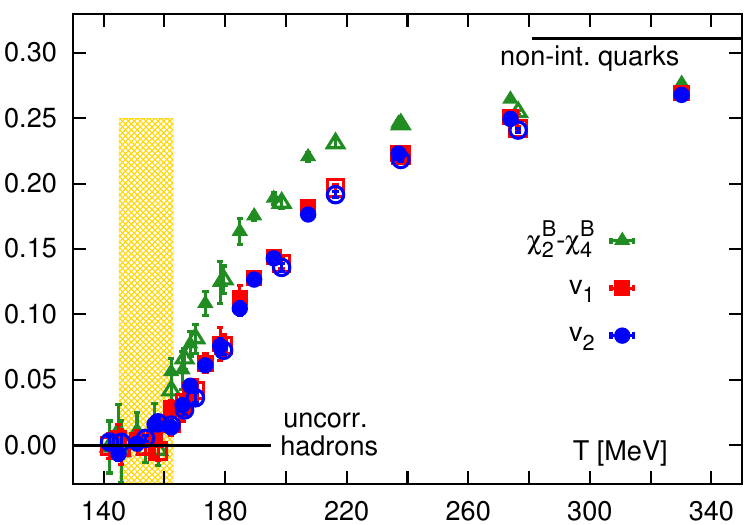}}
    \raisebox{-7pt}{\includegraphics[width=0.463\textwidth]{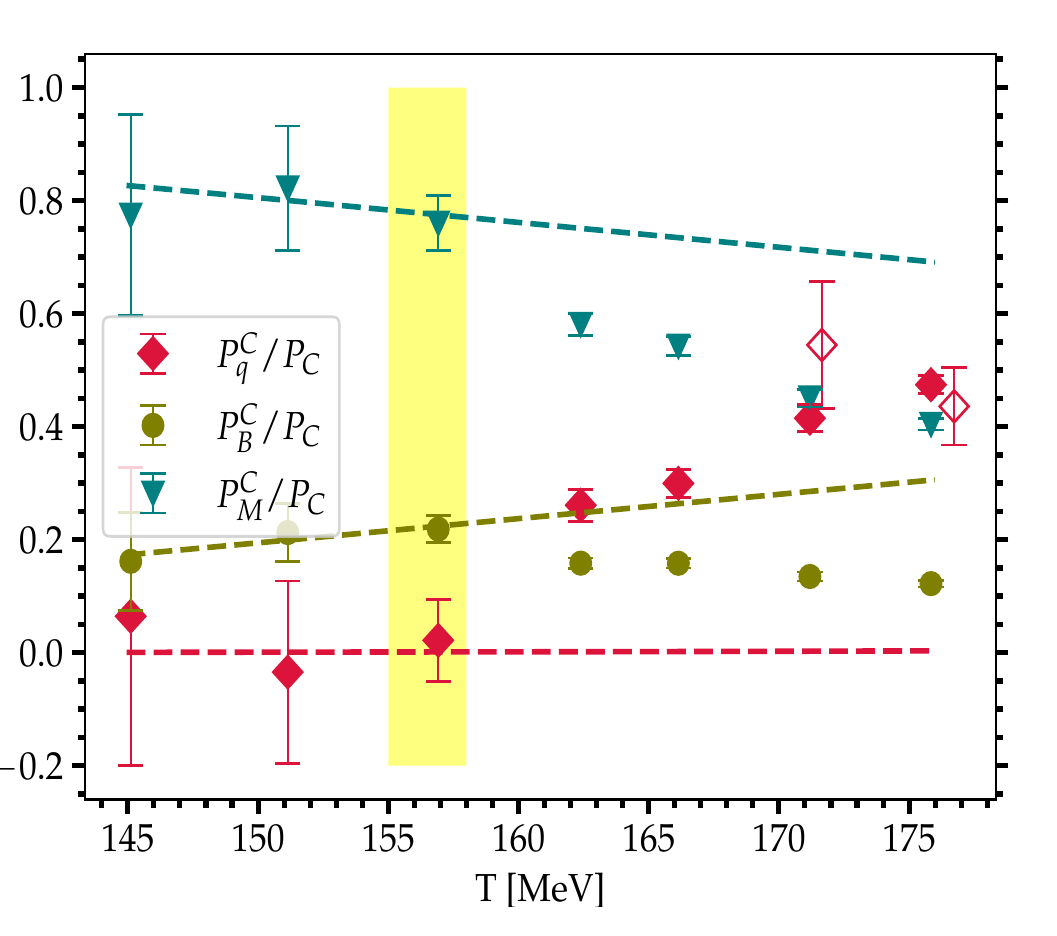}}
    \caption{Left: Two combinations, $v_1$ and $v_2$, of
strangeness fluctuations and baryon-strangeness correlations
that vanish identically if the system is described by an uncorrelated gas of hadrons. Also shown is the difference of
quadratic and quartic baryon number fluctuations, $\chi_2^B-\chi_4^B$, which has to vanish under the same condition. The figure is taken from \cite{Bazavov:2013dta}. Right: Partial pressures of charmed mesons, charmed baryons and charm quarks as functions of temperature. All three observables have been normalized to the total partial charm pressure. The dashed lines show corresponding results obtained from the QM-HRG model. Filled (open) symbols show the results for $N_\tau=8$ ($N_\tau=12$) lattices. The yellow band represents $T_{pc}$ with its uncertainty. The figure is taken from \cite{Bazavov:2023xzm}.}
    \label{fig:v1v2}
\end{figure}

In order to get some insight into the physics at nonzero density, one can formally expand the pressure in terms of chemical potentials. We introduce here four independent chemical potentials of of the hadronic net charges, i.e. baryon number, electric charge, strangeness and charm as $\mu_B, \mu_Q, \mu_S, \mu_C$. The expansion of the dimensionless pressure $p/T^4$ is given as 
\begin{equation}
    \frac{p}{T^4}=\sum_{i,j,k,l=0}^{\infty}\frac{1}{i!j!k!l!}\chi_{ijkl}^{BQSC}\hat\mu_B^i\hat\mu_Q^j\hat\mu_S^k\hat\mu_C^l\;,
    \label{eq:taylor}
\end{equation}
where $\hat\mu_X=\mu_X/T$, for $X\in\{B,Q,S,C\}$
The cumulants $\chi_{1,j,k,l}^{B,Q,S}$ are obtained as partial derivatives, defined as 
\begin{equation}
    \chi_{i,j,k,l}^{B,Q,S,C}=\left.\frac{\partial^{(i+j+k+l)}(p/T^4)}{\partial \hat\mu_B^i\hat\mu_Q^j\hat\mu_S^k\hat\mu_C^l}\right|_{\vec\mu=0}\;.
    \label{eq:cumulants}
\end{equation}
They are very useful quantities for at least three reasons: (i) they encode the partition function and thus provide access to the QCD phase diagram \cite{Bollweg:2022rps}, as we will discuss in the next section. (ii) They can also be measured in heavy ion experiments. Matching lattice QCD calculations with the experimental data, results in a model free  way to determine freeze-out parameter \cite{Bazavov:2012vg, Bazavov:2015zja}. (iii) They can be used to discuss the relevant degrees of freedom in the system and are thus sensitive to the deconfinement transition. This can be done, e.g., by disentangling different strangeness sectors \cite{Bazavov:2013dta}. As long as the ideal hadron resonance gas (HRG) is a good description of the system the pressure can be written as sum over the partial pressures
\begin{equation}
    (p/T^4)=(p^{|S|=0}_{M/B}/T^4)+(p^{|S|=1}_{M}/T^4)+(p^{|S|=1}_{B}/T^4)+(p^{|S|=2}_{B}/T^4)+(p^{|S|=3}_{B}/T^4)\;.
\end{equation}
Here $p_M,p_B$ denotes the mesonic, baryonic partial pressure, respectively. Using all strangeness fluctuations, and baryon strangeness correlations up to fourth order ($\chi_2^S, \chi_4^S, \chi_{11}^{BS}, \chi_{13}^{BS}, \chi_{22}^{BS, }\chi_{31}^{BS}$) will leave us with six equations for the four partial pressures with nonvanishing  strangeness \cite{Bazavov:2013dta}. The system is thus overconstrained and we can find two linear independent combinations, denoted as $v_1,v_2$, that should identically vanish as long as the HRG is a good description of the system. Results for these combinations are shown in Fig.~\ref{fig:v1v2} (left), obtained with (2+1)-flavor of HISQ fermions on $N_\tau=8$ lattices.
We find that below $T_{pc}$ the description in terms of an uncorrelated gas of hadrons is correct but quickly looses its validity for $T>T_{pc}$, i.e. the melting of strange hadrons starts just above $T_{pc}$. In particular, the agreement with the HRG is much improved in the vecinity of $T_{pc}$, once additional strange hadrons are added to the experimentally established list of hadrons \cite{Bazavov:2014xya}, which are predicted by quark model calculations \cite{Capstick:1986ter, Ebert:2009ub} and observed in
lattice QCD spectrum calculations \cite{Edwards:2012fx}. 

A similar analysis can be performed in terms of the charmed degrees of freedom \cite{Bazavov:2014yba, Bazavov:2023xzm, Sharma:2025zhe}, see Fig.~\ref{fig:v1v2} (right). Here the partial pressures of the $|C|=2$ and $|C|=3$ baryons are negligible since they are exponentially suppressed by their large masses. Current lattice simulations have not the precision to resolve these exponentially small contributions. On the other hand we can project onto the partial pressure of charm quarks as well. We find that the agreement with the HRG is good below $T_{pc}$, once additional charmed hadrons from quark model calculations \cite{Chen:2022asf, Ebert:2009ua, Ebert:2011kk} are added to the PDG list of experimentally established charmed hadrons. Again, the melting of charmed hadrons starts immediately above $T_{pc}$, where also charmed quarks appear as new degrees of freedom. Remarkably, already at $T>1.1 T_{pc}$ ($\gtrsim 175$ MeV) the partial pressure of charm quarks generates half of the total charm pressure. 

In addition we can also monitor above which temperature the conserved charge fluctuations become consistent with perturbative QCD calculations. Some of the cumulants, involving only one derivative of baryon number/electric charge, start to differ only at $\mathcal{O}(\alpha_s\ln(\alpha_s))$ from the free gas result \cite{Blaizot:2001vr},  with $\alpha_s$ being the strong coupling constant, and are thus in agreement with perturbative estimates already at $T\gtrsim 250$ MeV. However, in general we find that the cumulants become perturbative for $T\gtrsim 450-700$ MeV. In the temperature range between $T_{pc}<T<(2-3)T_{pc}$ we thus find a strongly interacting gas of color charges and excitations, sometimes also called "stringy liquid" \cite{Rohrhofer:2019qwq, Rohrhofer:2019qal}. This is the temperature range where an additional emerging chiral-spin symmetry of the color charges has been proposed \cite{Rohrhofer:2019qwq, Rohrhofer:2019qal, Glozman:2022lda}.

\section{The beam energy scan and the QCD critical point\label{sec:CEP}}
\begin{figure}
    \centering
    \includegraphics[width=0.565\textwidth]{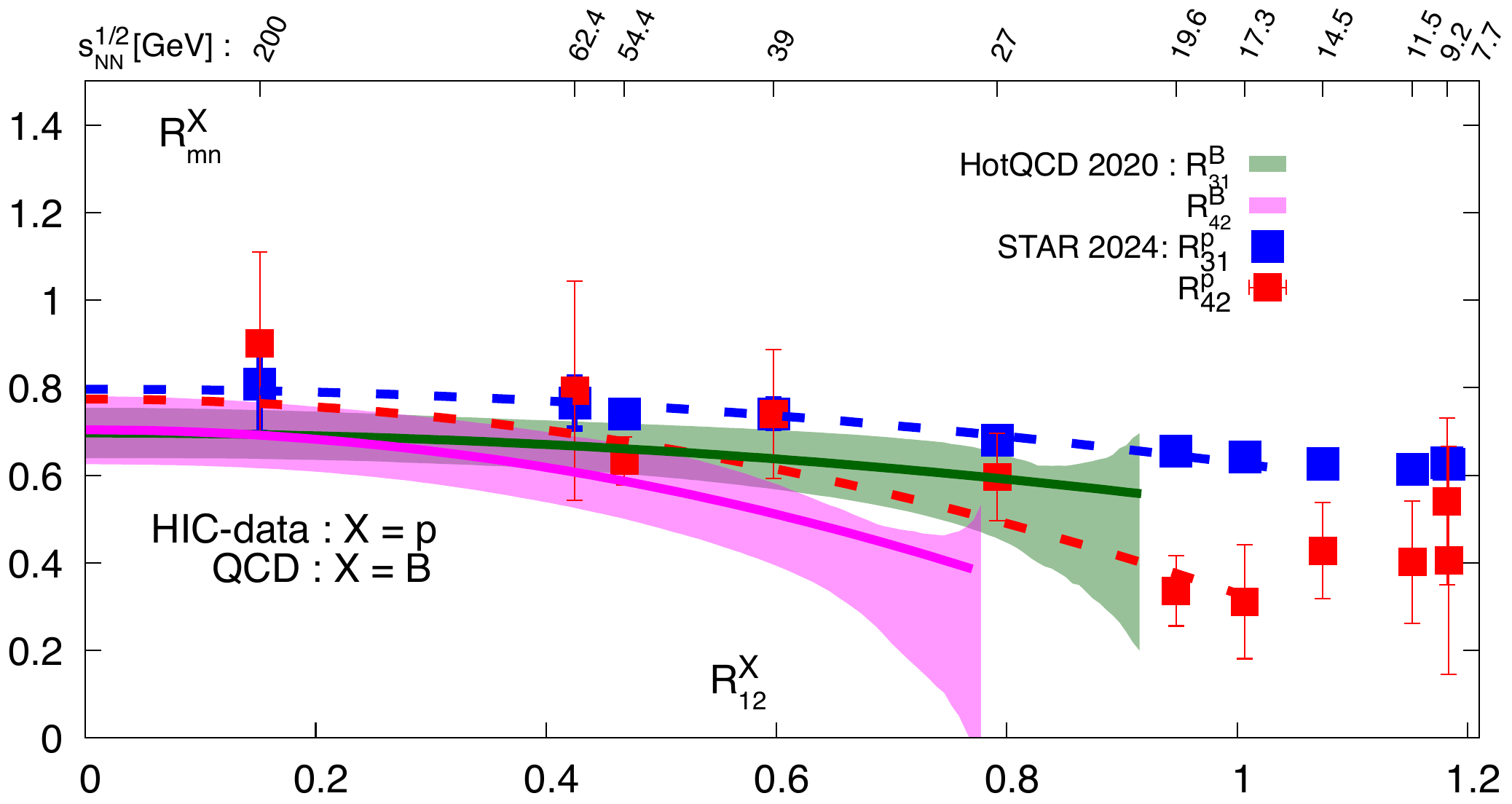}
    \includegraphics[width=0.415\textwidth]{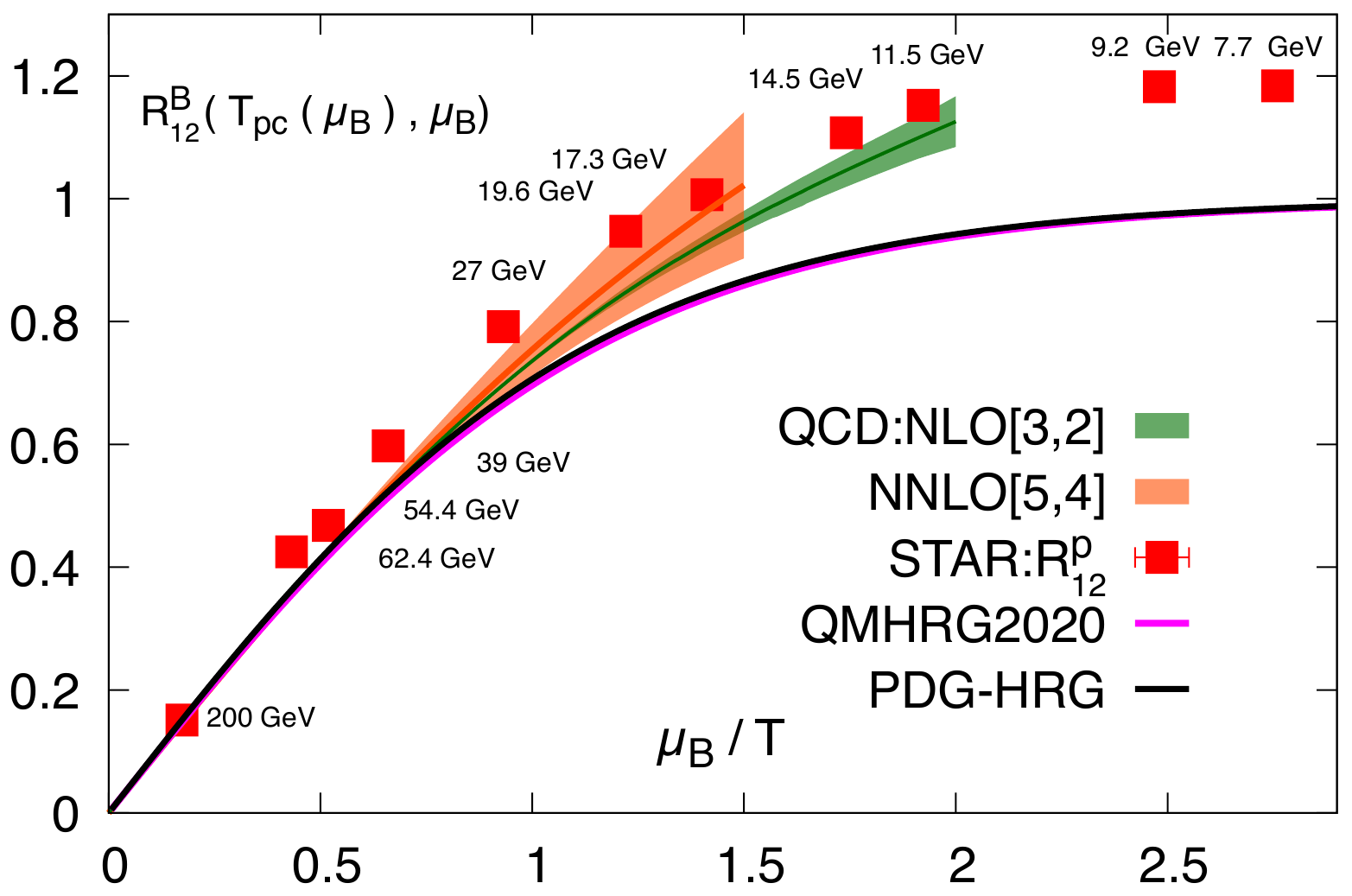}
    \caption{Left: Skewness ($R_{31}^X$) and kurtosis ($R_{42}^X$) ratios as function of $R_{12}^X$. Here $X=p,B$ for heavy ion and lattice data, respectively. Dashed lines are fits to the experimental data, bands are lattice QCD calculations from \cite{Bazavov:2020bjn}. Right: $R_{12}^X$ as function of $\mu_B/T$. The bands show NLO [3,2] and NNLO [5,4] Padé resummations of the Taylor coefficients calculated on the lattice. Also shown are HRG model calculations.}
    \label{fig:kurtosis}
\end{figure}

The cumulants of the conserved hadronic charges, as defined in Eq.~(\ref{eq:cumulants}), are accessible from the event-by-event distributions of the measured particle yields in heavy ion collisions. Usually we consider ratios of cumulants of charge $X$, we denote $R_{mm}^X=\chi_n^X/\chi_m^X$, in order to eliminate the leading dependence on the freeze-out volume. A non-monotonic behaviour in the kurtosis $R_{42}^B$ was proposed as signal for the critical point \cite{Stephanov:2011pb}. Recently the beam energy scan (BES) program II at the Relativistic Heavy Ion Collider (RHIC) was concluded. The collected data is expected to reduce statistical and systematic uncertainties over the BES-I results on the kurtosis, where non-monotonicity was found with a significance of 3.1 $\sigma$ \cite{STAR:2020tga}. Preliminary BES-II results have been presented \cite{Xu:2025kxa}. Unfortunately non-monotonic behavior could not be established so far, yet the general trend is consistent with the presence of a critical point \cite{Stephanov:2011pb}. It is also interesting to note that at low densities the data is well described by equilibrium thermodynamic calculations from lattice QCD as shown in Fig.~\ref{fig:kurtosis}.
To remove the model dependent extraction of the chemical potential $\mu_B/T$, associated with the collision energy $\sqrt{s_{NN}}$, we plot the skewness ($R_{31}^X$) and kurtosis ($R_{42}^X$) as function of the density $R_{12}^X$. 
Dashed lines correspond to fits to the experimental data, while bands are lattice QCD calculations from \cite{Bazavov:2020bjn}. Up to a small horizontal shift, the agreement between experimental measurements and lattice data is good for $R_{12}^X<0.9$. The shift indicates that the freeze-out temperature is below the crossover line, on which the lattice data is plotted. 

In order to push lattice QCD calculation towards chemical potential values of $\mu_B/T\lesssim 2.5$, coefficients in the Taylor series (Eq.~(\ref{eq:taylor})) need to be known up to the eighth order \cite{Bollweg:2022rps}. Unfortunately, errors on the eights order are still very large \cite{Borsanyi:2023wno}, due to cancellations between different contributions, which are as large as $\mathcal{O}(V^4)$, where $V$ is the volume. The Taylor series has a finite radius of convergence \cite{Giordano:2019slo, Giordano:2019gev, Pasztor:2020dur}. Estimates on the radius can be used to obtain bounds on the QCD critical point \cite{Bollweg:2022rps, Borsanyi:2025dyp}. To overcome the radius of convergence, various resummation schemes are discussed \cite{Bollweg:2022rps, Borsanyi:2021sxv,Dimopoulos:2021vrk, Mitra:2022vtf, Mitra:2023csk}. 

Recently, universal scaling related to the Lee-Yang edge singularity has been exploit to determine the location of the QCD critical point \cite{Dimopoulos:2021vrk, Clarke:2024ugt}. From a rational multi-point approximation of the baryon number density at imaginary chemical potentials, singularities in the complex $\mu_B$ plane are estimated. The multi-point Padé procedure naturally extents the Taylor expansion approach about $\mu_B=0$ \cite{Allton:2002zi} to include numerous expansion coefficients around multiple imaginary chemical potential values $\mu_B=i\theta_B$. The need for precise high-order coefficients is thus treated against multiple expansion points. Once a rational approximation of the baryon number density is known, the determination of its complex poles is strait forward. The closest singularity, which is found to be stable under the order of the Padé approximation, is identified with the Lee-Yang edge singularity. That this strategy is meaning full has been tested so far in the Ising Model \cite{Singh:2023bog}, the Roberge-Weiss transition in QCD \cite{Dimopoulos:2021vrk}, the 3-state Potts model and heavy-quark QCD \cite{Wada:2025ghc, Wada:2024qsk}. In all cases the expected universal scaling was found and known results on the location of the critical point could be reproduced. 

\begin{figure}
    \centering
    \includegraphics[width=0.48\textwidth]{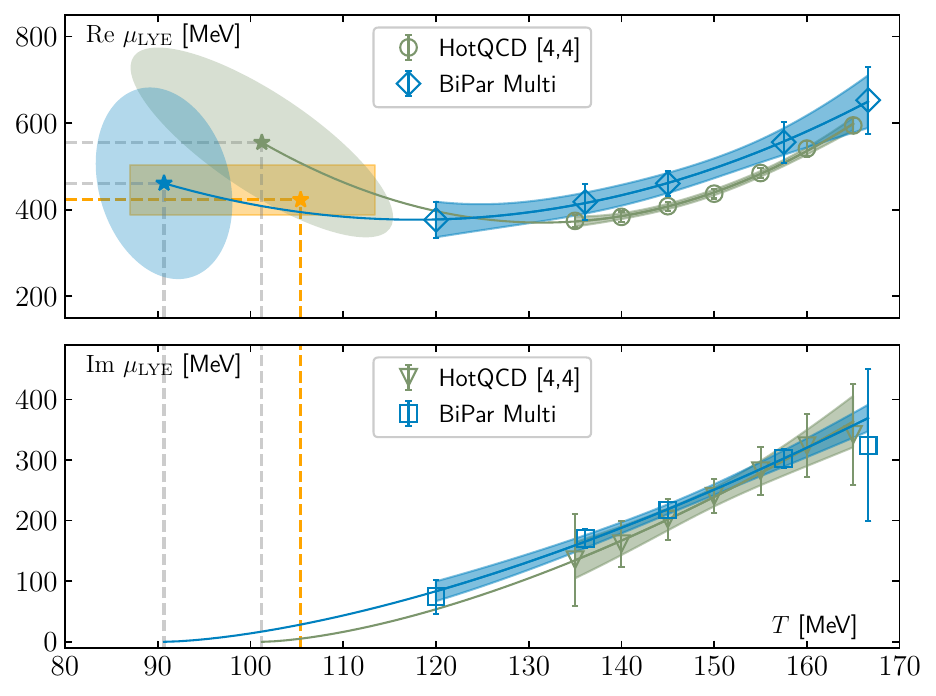}
    \includegraphics[width=0.50\textwidth]{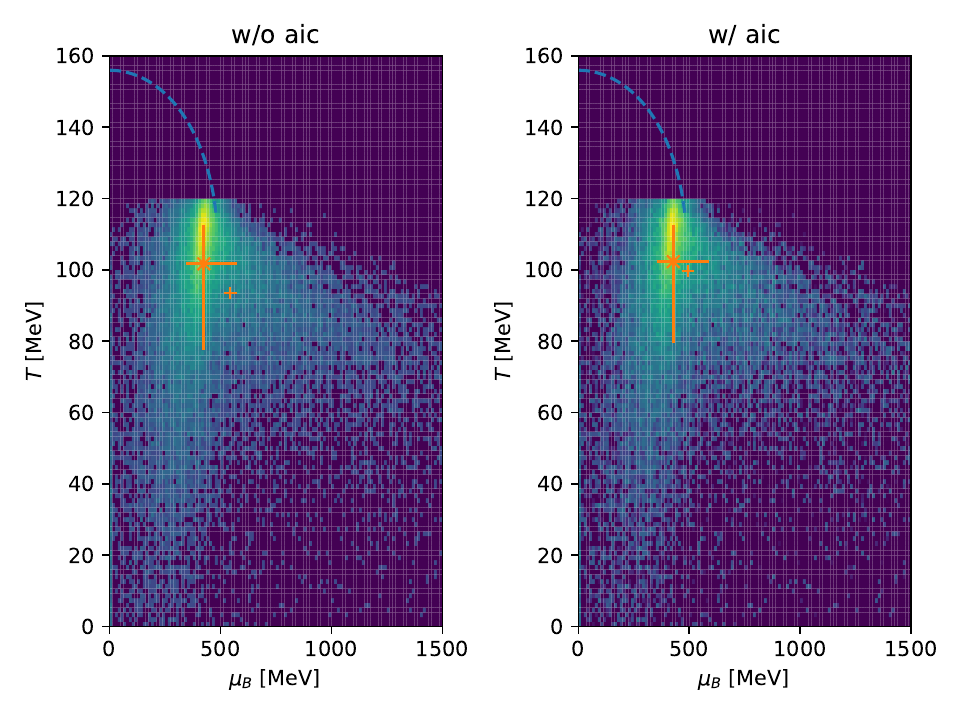}
    \caption{Left: Universal scaling fits to the temperature dependence of the poles in the complex $\mu_B$ plane. Data are from the multi-point Padé approach at imaginary chemical potential ($N_\tau=6$) \cite{Clarke:2024ugt}, and from the Taylor expansion approach about $\mu_B=0$ ($N_\tau=8$) \cite{Bollweg:2022rps}. Error ellipses are the 1-$\sigma$ confidence ellipses of the critical point  locations obtained from the fit, the orange bar denotes the weighted average over $\mathcal{O}(10^5)$ fits in the multi-point case. Right: histogram of all $\mathcal{O}(10^5)$ fit results in the multi-point case, weighted with and without Akaike information criterion. }
    \label{fig:CEP}
\end{figure}

Once the location of the Lee-Yange edge is known for some temperatures, they need to be extrapolated to the critical temperature, which is defined as the temperature where the imaginary part of the Lee-Yang edge vanishes. The scaling with temperature is fixed by the circle theorem of Lee and Yang \cite{Yang:1952be} and by assigning a universal position in the scaling variable $z$, i.e. $t/h^{1/\beta\delta}=z_{LY}$ \cite{Connelly:2020gwa}. Unfortunately, the scaling fields near the critical point are unknown. This is due to the fact that the reflection symmetry realized at the QCD critical point is an emergent symmetry and not manifest in the QCD Lagrangian. A frequently used mixing ansatz  \cite{Rehr:1973zz, Nonaka:2004pg} is given as
\begin{eqnarray}
t&=&\alpha_t(T-T_{cep})+\beta_t(\mu_B-\mu_{cep})\,, \nonumber \\
h&=&\alpha_h(T-T_{cep})+\beta_h(\mu_B-\mu_{cep})\,,
\label{eq:scaling_fields_CEP}
\end{eqnarray}
where $\alpha_t,\alpha_h, \beta_t$ and $\beta_h$ are the mixing parameters. 
One can easily verify that the ratio $-\beta_h/\alpha_h$ defines the slope of the first order line at the QCD critical point. With this ansatz and the assumption that the position of the Lee-Yang edge is universal, we can derive the temperature behaviour of the Lee-Yang edge in the complex chemical potential plane. 
We find that $\text{Im}[\mu_{LY}]
\sim(T-T_{cep})^{\beta\delta}$, for $T\searrow T_{cep}$, while $\text{Re}[\mu_{LY}]=\mu_{cep}-\beta_h/\alpha_h(T-T_{cep})+\mathcal{O}(T^2)$ \cite{Stephanov:2006dn}. 
A fit to the Lee-Yang scaling is shown in Fig.~\ref{fig:CEP} (left). The poles are obtained from the multi-point analysis at imaginary chemical potential ($N_\tau=6$) \cite{Clarke:2024ugt}, where [3,3]-, [4,4]- and [5,5]-Padés have been used, and from the eighth order Taylor expansion of the pressure about $\mu_B=8$ ($N_\tau=8$), which is resummed in a [4,4]-Padé. 
The corresponding estimates are shown in the upper panel, where the error ellipses are from the fit, while the yellow box indicates the weighted average over $\mathcal{O}(10^5)$ fits to the multi-point Padé data, that differ in the order of the Padé and the fit range. In Fig.~\ref{fig:CEP} (right) a two dimensional histogram of all critical point locations are shown, weighted with and without Akaike information criterion.
\begin{wrapfigure}{r}{0.6\textwidth}
     \centering
    \includegraphics[width=0.6\textwidth]{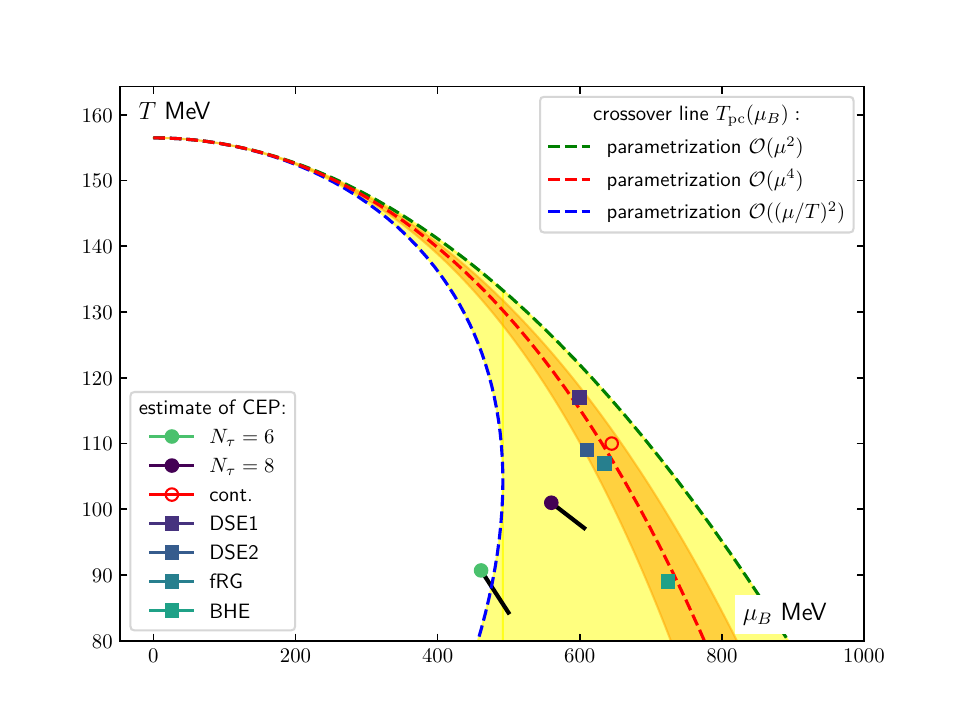}
    \caption{compilation of critical point locations from lattice QCD, FRG, DSE and BHE. Also shown are second and fourth order parametrizations of the crossover line.}
    \label{fig:crossover}
\end{wrapfigure}
For the location of the CEP we find $T_{cep}=105^{+8}_{-18}$ MeV, and $\mu_{cep}= 422^{+80}_{-35}$ MeV. Similar scaling fits for the data from \cite{Bollweg:2022rps}, combined with a conformal mapping, has been presented \cite{Basar:2023nkp}. Preliminary results on Lee-Yang edge scaling applied to the Budapest-Wuppertal data has been discussed \cite{Adam:2025pii}. 

We are now in the situation where estimates of the CEP location from different methods are available, including Functional Renormalization group calculations (FRG) \cite{Fu:2019hdw}, a generalized FRG approach \cite{Gao:2020fbl}, Dyson-Schwinger (DSE) equations \cite{Gunkel:2021oya} and Black Hole engineering (BHE) \cite{Critelli:2017oub, Hippert:2023bel}. They all seem to cluster in a narrow range of the QCD phase diagram, favouring $T_{cep}\approx 110$ MeV and $\mu_{cep}\approx 600$ MeV, see Fig.~\ref{fig:crossover}. A naive continuum estimate from Lee-Yang scaling in QCD, based on $N_\tau=6,8$ \cite{Clarke:2024ugt} falls also in that region. These estimates are also consisting with the curvature estimates for the crossover line from lattice QCD, albeit a second order estimate of the crossover line is already ambiguous above $\mu_B>400$. They are also consistent with the constraints obtained from the estimated convergence radius \cite{Bollweg:2022rps} and improved equation of state \cite{Borsanyi:2025dyp}. 

\section{Summary and outlook} 
The structure of the QCD phase diagram is largely determined by chiral symmetry breaking and deconfinement. These two phenomena are intimately intertwined and in particular the latter is not yet fully understood. The influence of various different control parameter on chiral and deconfinement transitions is discussed in a increasingly quantitative manor and the phase diagram is understood as a multi-dimensional object. Important control parameter are the quark masses, chemical potentials, the external magnetic fields and number of flavors, motivated by phenomenological considerations in cosmology, astro and heavy ion physics. 

In the vicinity of a second order transition the phenomena of universal critical scaling is expected and has been observed in lattice QCD data of the chiral order parameter. Applying knowledge from universal scaling can help to further refine our understanding of the phase diagram and to determine none universal parameter such as the position of the phase transition. However, current computations also enter the precision to check the universality class of the phase transition. Universal scaling in the vicinity of the QCD critical point can also be discussed in terms of the universal position of the Lee-Yang edge singularity. Recent lattice estimates on the QCD critical point are based on this method and are located in the same region of the QCD phase diagram as Functional Renormalization Group (FRG), Dyson-Schwinger (DSE), and Black Hole Engineering (BHE) calculations of the critical point location, i.e. at $T_{cep}\approx 110$ MeV and $\mu_{cep}\approx 600$ MeV.

As the beam energy scan program II at RHIC has just been concluded and a smoking gun signal of the QCD critical point is not been detected, we are now awaiting first beams with the Compressed Baryonic Matter (CBM) detector at the Facility for Anti-proton and Ion Research (FAIR), which are expected in 2028 \cite{Hohne:2024kpd}. 

\section*{Acknowledgements}
CS acknowledges all members of the HotQCD and Bielefeld-Parma Collaboration for valuable discussions. This work was supported (i) by the European Union’s Horizon 2020 research and innovation program under the Marie Sklodowska-Curie Grant Agreement No. H2020-MSCAITN-2018-813942 (EuroPLEx), (ii) by The Deutsche Forschungsgemeinschaft (DFG, German Research Foundation) - Project No. 315477589-TRR 211 and the PUNCH4NFDI consortium supported by the Deutsche Forschungsgemeinschaft (DFG, German Research Foundation) with grand 460248186 (PUNCH4NFDI).

\providecommand{\href}[2]{#2}\begingroup\raggedright\endgroup

\end{document}